\documentclass[12pt]{article} 

\usepackage{osajnl2}
\usepackage[draft]{hyperref}
\usepackage{amsmath}

\begin{document}


\title{Observation of time correlation function of multimode two-photon pairs on a rubidium D$_2$ line }


\author{Fu-Yuan Wang, Bao-Sen Shi$^*$, and Guang-Can Guo}

\address{
Key Laboratory of Quantum Information,University of Science and Technology of China,Hefei 230026,China 
\\
$^*$Corresponding author: drshi@ustc.edu.cn
}

\begin{abstract} We report the  generation of a type-I multimode two-photon
  state on a rubidium D$_2$ line (780 nm) using periodically poled
  KTiOPO$_4$ crystals. With a degenerate
  optical parametric oscillator far below threshold
, we observe an oscillatory correlation function, the
  cross-correlation between two photons shows a cavity bandwidth of about
  7.8 MHz. We also use a  Fabry-P$\acute{\rm{e}}$rot etalon to filter its
  most longitudinal modes and observe its time correlation function. The
  experimental data are well fitted to  theoretical curves. This system could
  be utilized for demonstrating storage and retrieval of narrowband  
  photons in Rb atomic ensembles, which is important for long-distance quantum
  communication.
\end{abstract}


 
 \noindent Sources for creating entangled photon pairs are essential parts of many
quantum information protocols and quantum optical experiments, such as quantum cryptography\cite{PhysRevLett.67.661},
teleportation\cite{PhysRevLett.70.1895,Bouwmeester575}, dense
coding\cite{PhysRevLett.76.4656},
quantum computation\cite{deutsch1992} and others. To date, the most widely used way of
obtaining the entangled photon pairs is the spontaneous parametric
down-conversion (SPDC)\cite{PhysRevLett.25.84}  in a nonlinear crystal. In quantum information
field, quantum memory must be used in order to realize the long-distance
quantum communication, therefore transferring quantum state between the light
and the memory is the key point for realizing such long-distance
communication. Recently, there are many primitive experiments in this
field\cite{nature01714,Wal07112003,nature04353}, in which, an atomic system
like a hot atomic vapor cell or a cold atomic
cloud of Rb or Cs, is used as a quantum memory, and a photon is used as a
flying bit. The key point to realize the efficient coupling between the atom
and the photon is how to get a photon which has a comparable bandwidth with
the natural bandwidth of the atom. This requirement excludes sources such as molecules in
solid-state matrices\cite{35035032}, nitrogen-vacancy centers at room
temperature\cite{VincentJacques02162007} and
quantum dots\cite{nature01086}.There are some possible methods for generating a narrow band
photon: one is base on the spontaneous Raman scattering in an atomic system,
which has been used in schemes shown in
Refs.\cite{nature01714,Wal07112003,nature04353}, but the experimental
effort is high. Another one is based on SPDC with a nonlinear crystal in a cavity, as been demonstrated
in Ref.\cite{PhysRevLett.83.2556}. Following this work, several
groups\cite{PhysRevA.70.043804,kuklewicz:223601,scholz:191104} recently 
performed the almost same experiments as in Ref.\cite{PhysRevLett.83.2556}. Our group is now involving in the realization of
the quantum memory based on a cold Rb atomic ensemble (D$_2$ line at 780 nm)
trapped in a magneto-optical trap. We plan to transfer the quantum state
between an atomic ensemble and a narrow-band single photon, which we hope to
prepare by the SPDC in a ring cavity. In this work, we report the 
experimental generation of a multimode two-photon state at 780 nm via a
degenerate optical parametric oscillator (OPO) far below threshold. This
wavelength corresponds to the D$_2$ line of Rb atom. We have
observed the oscillatory correlation of multimode two-photon pairs. This is, to our best
knowledge, the first time to get this kind oscillatory correlation function 
at Rb D$_2$ line. Cross-correlation  between
two photons shows a cavity bandwidth of 7.8 MHz. Besides, our experiment is
different from others: in Ref.\cite{PhysRevLett.83.2556}, a bigger and more expensive Ti: Sapphire
laser is used as the fundamental light for the second harmonic generation in a
cavity. Our fundamental source is a small and relatively cheap
grating-stabilized external cavity diode laser (ECDL) (Toptica DL100), and its
frequency is locked to the transition line D$_2$ of Rb$^{87}$ atom (780 nm) by the
saturated absorption technique. In Ref.\cite{PhysRevA.70.043804}, a pair of cw type-I down
converters, one rotated by 90$^{\circ}$, in a ring cavity to produce
polarization-entangled photons. In this experiment, Ti: Sapphire laser and
KbNO$_3$ are used. In our experiment, a periodically poled KTiOPO$_4$
(PPKTP), which has larger nonlinear coefficient, is
used to compensate the relative lower power of the diode laser. In
Ref.\cite{kuklewicz:223601}, a cw type-II down converter is used to get a narrow-band photon pair
at 795 nm with an ECDL at 397.5 nm. In our experiment,
the pump laser at 390 nm for OPO is from frequency doubling of the 780 nm
laser in a cavity. In our experiment, a double resonant OPO is used. In Ref.\cite{scholz:191104}, a single
resonant OPO is used to prepare a narrow-band photon. The OPO is locked
by Hansch-Couillaud method. Furthermore, we use a Fabry-P$\acute{\rm{e}}$rot
etalon to filter most longitudinal modes of the photons and observe its time
correlation function. The results show that there is no oscillatory
correlation in this case.
\\
\indent  The theory of the output from an OPO
 far below threshold has been discussed in
 Ref.\cite{PhysRevA.62.033804,Gardiner}. Since the large bandwidth of PDC
 photons, there are numerous nondegenerate conjugate pairs together with
 degenerate pairs. The nondegenerate pairs are located at two sides of
 degenerate frequency of OPO($\omega_0$) with a spacing of
 $\Delta\Omega_{opo}$($\Delta\Omega_{opo}$ is the free spectrum range of the
 OPO cavity).  The output operator is given by the
 following equation($\omega_{\pm m}=\omega_0\pm
 m\Delta\Omega_{opo},m=0,1,2\cdots $)\cite{PhysRevA.62.033804,Gardiner}:\begin{align}
   a_{out}(\omega+\omega_m)&=G_1(\omega)a_{in}(\omega_m+\omega)+g_1(\omega)a_{in}^{\dag}(\omega_{-m}-\omega)\nonumber\\
    &+G_2(\omega)b_{in}(\omega_m+\omega)+g_2(\omega)b_{in}^{\dag}(\omega_{-m}-\omega),\label{output}
 \end{align}

\noindent with \begin{align}
   G_1(\omega)=\frac{\gamma_1-\gamma_2+2i\omega}{\gamma_1+\gamma_2-2i\omega},\
   \ g_1(\omega)=\frac{4\epsilon\gamma_1}{(\gamma_1+\gamma_2-2i\omega)^2},\nonumber\\
   G_2(\omega)=\frac{2\sqrt{\gamma_1\gamma_2}}{\gamma_1+\gamma_2-2i\omega},\
   \ g_2(\omega)=\frac{4\epsilon\sqrt{\gamma_1\gamma_2}}{(\gamma_1+\gamma_2-2i\omega)^2},\label{output2}
 \end{align}
Here $a_{in}(\omega)$ and $b_{in}(\omega)$ represent the vacuum mode entering
 the OPO cavity and the unwanted vacuum mode coupled-in due to losses in the
 system, respectively. $\gamma_1$ and $\gamma_2$ are the coupling
 constants for $a_{in}$ and $b_{in}$, respectively. $\epsilon$ is the
 single pass parametric amplitude gain.  The intensity correlation
 function of two down-converted photons is defined as\begin{eqnarray}
   \Gamma^{(2)}(\tau)= \langle
   \hat{E}^{(-)}(t)\hat{E}^{(-)}(t+\tau)\hat{E}^{(+)}(t+\tau)\hat{E}^{(+)}(t)\rangle. \label{intfun}
 \end{eqnarray}
with\begin{eqnarray}
  \hat{E}^{(+)}(t)=[\hat{E}^{(-)}(t)]^\dag=\frac{1}{\sqrt{2\pi}}\int
  d\omega\hat{a}(\omega)e^{-i\omega t}.\label{eqs4}
\end{eqnarray}
From Eqs. (\ref{output}),(\ref{output2}),(\ref{intfun}), and 
(\ref{eqs4}) and with some calculations one can find
that\cite{PhysRevA.68.015803}\begin{align}
  \Gamma^{(2)}(\tau)=&|\epsilon|^2 \left(\frac{F}{F_0}\right)^2
  \Bigg[\left(\frac{2|\epsilon|(2N+1)}{\Delta\omega_{opo}}\right)^2\nonumber\\
 & +e^{-\Delta\omega_{opo}|\tau|}\frac{\sin
  ^2 [(2N+1)\Delta\Omega_{opo}\tau/2]}{\sin
  ^2(\Delta\Omega_{opo}\tau/2)}\Bigg]. \label{eqs5}
\end{align}
Here $\Delta\omega_{opo}$ is the bandwidth of the OPO. $\tau$
represents the delay time. (2N+1) represents the number of modes of the
output optical fields from OPO. $F$ and $F_0$ is the
finesse of the OPO with and without loss, respectively. The detail 
explanation of this formula can be shown in
Ref.\cite{PhysRevA.68.015803}.  One of the key points to observe the
oscillatory correlation function is that
$\tau_{opo}=2\pi/\Delta\omega_{opo}$, which is round-trip time of the OPO
cavity, must larger than the resolving time of detectors, $\tau_D$. In our
first
experiment, $\tau_D$ and $\tau_{opo}$ are 220 ps and 1.63 ns,
respectively, therefore $\tau_D<\tau_{opo}$ in our experiment. 
\\
\begin{figure}
\begin{center}
\includegraphics[width=0.5\textwidth]{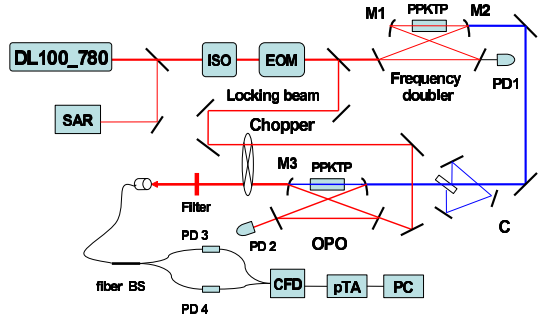}
\end{center}
\caption{Experiment setup. DL100\_780, an external-cavity diode laser
  operated at 780nm; ISO, optical isolator;EOM, electro-optic modulator; OPO,
  far below threshold degenerate optical parametric oscillator; SAR, saturated
  absorption resonator; PD1 and PD2, fast photodetector for cavity locking; PD3 and
  PD4, avalanche photodetectors; CFD, constant-fraction discriminator; pTA,
  picosecond time analyzer.}
\label{fig:setup}
\end{figure}

 In this paper, we obtain multimode
 narrowband two-photons with a PPKTP crystal via an OPO far below
 threshold, and observe the oscillatory correlation function of
 multimode two-photon pairs. That, to the best of our
 knowledge, is the first time to get the oscillatory correlation
 function at Rb D$_2$ line. A schematic drawing of the experiment set-up is
 shown in Figure ~\ref{fig:setup}. A cw ECDL with 780 nm wavelength is used to produce UV light
at 390 nm via a ring cavity. A 10 mm long type-I phase matched PPKTP crystal
with a domain period of 2.95 $\mu\mathrm{m}$  is used as the doubler. The frequency of the
laser is locked to the transition line D$_2$ of Rb$^{87}$ atom (780 nm) by the
saturated absorption technique. The frequency of the doubling cavity is locked
to laser frequency by PDH method\cite{PDHmethod}. Please refer to Ref.\cite{wangoptcomm} for
detail about frequency doubling.  The OPO cavity is composed of two concave mirrors of
curvature radius 80 mm and two plane ones. The concave mirror(M3) has partial
transmission about 5\% at 780nm and works as an output coupler, while the others
are high  reflectance mirrors. The round-trip length of the cavity is 480 mm
correspond to 0.625 GHz free spectral range(FSR) and 1.6 ns $\tau_{opo}$. The
distance between two spherical mirrors is about 100 mm, and a 10 mm long PPKTP
is placed between two spherical mirrors resulted in a waist of 40 $\mu$m
inside the crystal. The temperature of the crystal is controlled 
by a home-made thermoelectric cooler with
 the stability of $0.02\,^{\circ}\mathrm{C}$. The OPO cavity is locked to
 laser frequency by PDH method\cite{PDHmethod}. A chopper is used to  cut the photons of locking
light reflected from the surface of the crystal to avoid possible background
noise. The colored glass filer is used to
cut the UV light. The triangle cavity is used to get
mode-matched between two
bow-tie-type ring cavities. The outputs from OPO are coupled into a 50/50 fiber
beam-splitter(NEWPORT P22S780BB50). The outputs of the beamsplitter are input
to single photon detectors(PekinElmer SPCM-AQR-14-FC). The outputs of 
the detectors are sent to a coincident circuit for coincidence
counting, which 
mainly consist of a picosecond time analyzer(ORTEC, pTA9308) and a
computer. 
\\
\begin{figure}
\begin{center}
\includegraphics[width=0.5\textwidth]{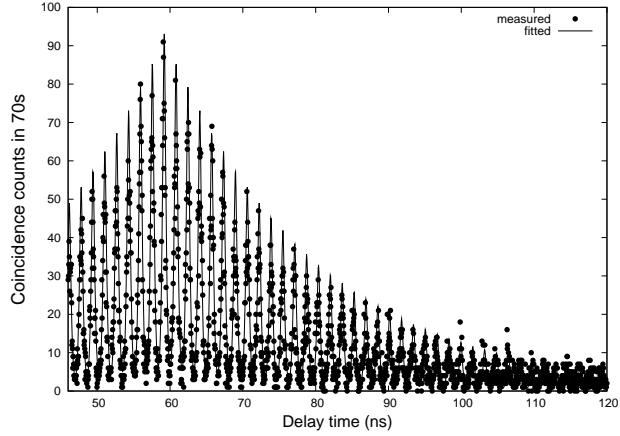}
\end{center}
\caption{Observed the oscillatory correlation function of multimode
  two-photon pairs. The coincidence counts are accumulated in 70 seconds. The
  pump power is about 15 $\mu$W. The fitted
  parameters of curve line are as follow: C$_1$=93, C$_2$=0,
$\Delta\omega_{opo}/2\pi$=7.8 MHz, $\tau_0$=59 ns,
$\tau_{opo}$=1.63 ns, $\tau_D$=220 ps.}
\label{fig:data}
\end{figure}
\indent Figure ~\ref{fig:data} shows the coincidence counts of multimode two-photons
at about 15 $\mu\textrm{W}$ pumping power. The points are the measured
data, and the line is the fitted curve using Eq.
({\ref{gamma}})\cite{PhysRevA.68.015803}\begin{align}
  \Gamma^{(2)}_c(\tau)=& C_1\Biggl[C_2+e^{-\Delta\omega_{opo}|\tau-\tau_0|}\nonumber\\
  &\times\sum_n\left(1+\frac{2|\tau-n\tau_{opo}-\tau_0|\ln2}{\tau_D}\right)\nonumber\\
  &\times
  \exp \left(-\frac{2|\tau-n\tau_{opo}-\tau_0|\ln2}{\tau_D}\right)\Biggr],\label{gamma}
\end{align}
Considering the probability distribution of the timing jitter of
detectors($p(\tau)$), Eq. (\ref{eqs5}) needs to be averaged over $p(\tau)$,
which is assumed to be proportional to $\exp(-2|\tau|\ln/\tau_D)$ here. We sum
all the modes from n=$-N$ to n=$N$. C$_1$ and C$_2$ are constant, which
represent the coincidence counts from the background and that from the
different photon pairs, respectively. The results of the simulation are as
follow: C$_1$=93, C$_2$=0,
$\Delta\omega_{opo}/2\pi$=7.8 MHz, $\tau_0$=59 ns,
$\tau_{opo}$=1.63 ns, $\tau_D$=220 ps. The data is accumulated in 70 seconds  with 4.88 ps time bins revolution.  
The theoretical curve fits the experiment data quite well. The pTA has a range
of dead time at the beginning of each scan, which is about 45 ns. So the left part of this
comb-like wavepacket isn't included in Figure ~\ref{fig:data}. 
\\
\begin{figure}[h]
  \begin{center}
	\includegraphics[width=0.5\textwidth]{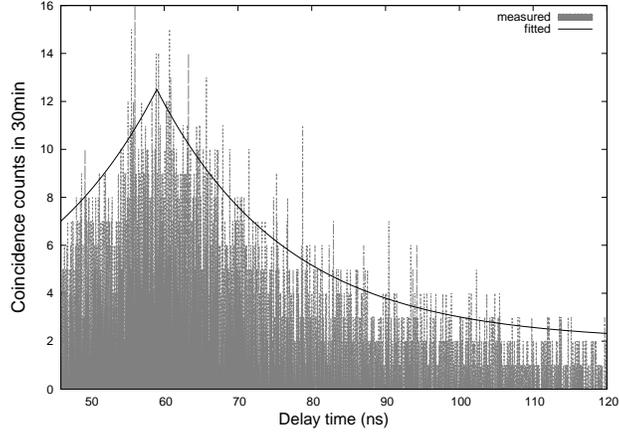}
  \end{center}
  \caption{The data shown in this figure are accumulated in 30 minutes. The pump
  power is about 90 $\mu$W. There is no obvious or regular oscillatory
   time correlation function.}
  \label{fig:smode}
\end{figure}
\indent We also put a Fabry-P$\acute{\rm{e}}$rot(FP) etalon with 13 GHz FSR and 1
GHz bandwidth after the out port of 
the OPO to filter most longitudinal modes of two-photon state. According to
our 0.625 GHz $\Delta\Omega_{opo}$, only less 10\% longitudinal modes can be
preserved. In this case, $\tau_{opo}$ is about 77 ps and is more less than $\tau_D$.
So the oscillatory of time correlation function like Fig. \ref{fig:data} will
disappear. In the experiment, we use a temperature controller to finely turn
the etalon's length, in order to let the degenerate frequency of
OPO($\omega_0$) have a maximum transmission.   The result is shown in Fig. \ref{fig:smode}, from which we could
find there is no obvious or regular oscillatory time correlation function.
Another thing we want to mention is that there are many small unregular peaks on the wavepacket,
for the two-photon state after the etalon is not perfect single mode.
\\
\indent In conclusion, we observed the oscillatory correlation function of  multimode
  two-photons at 780nm, which corresponds to D$_2$ transition of Rb
  atoms. The fitting line from Eq. (\ref{gamma}) almost overlaps with the
  experimental data. These results clearly show the good correlation between the two photons in a pair. The bandwidth of the photons is about 7.8 MHz,
which could make the efficient coupling between the photons and Rb atoms
possibly. We also use a FP etalon to filter its most longitudinal modes and
observe its time correlation function.
\\
\indent We thank Dr. Y. Zhang and Prof. Z-y Ou for their valuable helps
and Mr. J. S. Xu and Dr. C. F. Li for their kindly lending of pTA.
This work is supported by National Natural Foundation of Science, (Grant
No. 10674126), National Fundamental Research Program (Grant No.
2006CB921900), the Innovation fund from CAS, Program for NCET.


\end{document}